\documentclass[pss,a4paper]{sa}
\usepackage{times}
\usepackage{sa}
\usepackage[latin1]{inputenc}
\usepackage[T1]{fontenc}
\usepackage{amssymb,amsmath}
\usepackage{color}
\usepackage[]{graphicx}

\usepackage{graphicx}
\usepackage[mathscr]{eucal}
\usepackage[latin1]{inputenc}
\usepackage{amsmath}
\usepackage{amsfonts}
\usepackage{amssymb}
\usepackage{graphicx}

\usepackage{graphicx,amsmath,hyperref}
\usepackage{tikz}

\def\f{\frac}

\newcommand{\bce}{\begin{center}}
\newcommand{\ece}{\end{center}}
\newcommand{\be}{\nopagebreak[3]\begin{equation}}
\newcommand{\ee}{\end{equation}}
\newcommand{\ba}{\nopagebreak[3]\begin{eqnarray}}
\newcommand{\ea}{\end{eqnarray}}

\newcommand{\Feyn}[1]{#1\kern-0.65em/}
\definecolor{celeste}{rgb}{.23,.5,.7}

\def\bc{\begin{center}}
\def\ec{\end{center}}

\def\rg{\rangle}
\def\lg{\langle}

\def\de{\delta}

\def\la{\lambda}
\def\vl{\vec{\la}}

\def\de{\delta}
\def\la{\lambda}
\def\vl{\vec{\la}}       
\usepackage{wrapfig}
\begin{document}
%%    The information for the title page will be placed between
%%    \begin{document} and \maketitle. The order of most entries
%%    is determined by the class file and can not be changed by
%%    rearranging them. The maketitle command follows after the
%%    abstract.
%%
%%    Commands from here to \title will be completed by the publisher. Please do not make changes.
%%
\renewcommand{\copyrightyear}{2011}
\DOIsuffix{theDOIsuffix}
%%
%% issueinfo for header and copyright line
\Volume{4} \Issue{1} \Copyrightissue{01} \Month{01} \Year{2006} %%
%%    First and last pagenumber of the article. If the option
%%    'autolastpage' is set (default) the second argument may be left empty.
\pagespan{1}{}
%%
%%    Dates will be filled in by the publisher. The 'reviseddate' and
%%    'dateposted' (Published online) entry may be left empty.
$$\Receiveddate{\sf zzz} \Reviseddate{\sf zzz} \Accepteddate{\sf zzz}
$$\Dateposted{\sf zzz} %%
%%

%% \pretitle{Editor's Choice}

%% We have a short and a long form for the title. The short form
%% (optional argument) goes into the running head.
%\title[Manuscript preparation guidelines]{Manuscript preparation guidelines
%for authors of \\ Scientifica Acta in 2007/08}

\definecolor{lightgray}{gray}{0.85}

\hspace{368px}\colorbox{lightgray}{\color{black} \large{Physics}}\\

\title[Spinfoam Cosmology]{General covariant transition amplitudes in quantum cosmology}

\author{Francesca Vidotto}

%\email{vidotto@cpt.univ-mrs.fr}

\address{Centre de Physique Th\'eorique de Luminy\footnote{Unit\'e mixte de recherche (UMR 6207) du CNRS et des Universit\'es de Provence (Aix-Marseille I), de la M\'editerran\'ee (Aix-Marseille II) et du Sud (Toulon-Var); laboratoire affili\'e \`a la FRUMAM (FR 2291).\vskip2pt}
        , Case 907, F-13288 Marseille, EU;}
\address{Dipartimento di Fisica Nucleare e Teorica, Universit\`a degli Studi di Pavia, and}
\address{ Istituto Nazionale di Fisica Nucleare, Sezione di Pavia, via A. Bassi 6, I-27100 Pavia, EU.}

%% Information for the first author.
%\author[A. Bendiscioli]{A. Bendiscioli\inst{1}}
%%
%% Information for the second author
%\author[M. Belotti]{M. Belotti\inst{1,2}}
%%
%% Information for the third author
% ..........
%%
%\address[\inst{1}]{Biblioteca interdipartimentale di Fisica, Università di Pavia, Italy \\adsad@fhgfgh.it}
%\address[\inst{2}]{Dipartimento di Fisica A. Volta, Università di Pavia, Italy \\tsdr@ghdft.it }
%%
%    \dedicatory{This is a dedicatory.}

%% maketitle must follow the abstract.
\maketitle                   % Produces the title.

%% If there is not enough space inside the running head
%% for all authors including the title you may provide
%% the leftmark in one of the following three forms:

%% \renewcommand{\leftmark}
%% {First Author: A Short Title}

%% \renewcommand{\leftmark}
%% {First Author and Second Author: A Short Title}

% not modify the following line
\renewcommand{\leftmark}
{F. Vidotto: General covariant transition amplitudes in quantum cosmology.}

%\title{Spinfoam Cosmology
%{\bfseries\itshape\selectfont quantum cosmology from the full theory}}

\begin{abstract}
The path-integral approach to %quantum 
cosmology consists in the computation of transition amplitudes between states of the quantum geometry of the universe. In the past, the concrete computation of these transitions amplitudes has been performed in a perturbative regime, breaking the full general covariance of the theory. Here I present how it is possible to define a general covariant path integral in quantum cosmology, by relying on the most recent results of the canonical and covariant formulations of Loop Quantum Gravity.
 I present two strategies that have been implemented.
The first starts from the full Spinfoam theory, i.e. the path-integral framework for Loop Quantum Gravity, and defines a cosmological system. This is not obtained from symmetry reduced variables that are successively quantized, but directly considering the approximations that are characteristic of the full theory. The Spinfoam Cosmology obtained in this way includes quantum fluctuations beyond standard perturbation theory.
The second strategy exploits  the Hamiltonian constraint of Loop Quantum Cosmology, that is exponentiated in the formal expression of the usual path integral. The result is a general covariant path integral, that reproduces the form of the amplitude in the full Spinfoam theory. Therefore, this procedure connects the canonical and the covariant formalisms.

\end{abstract}

%\section{Introduction}
\vspace{15mm}
Hamiltonian and path integral quantum mechanics are based respectively on the canonical and the covariant quantizations. The two frameworks are complementary and the relation connecting one to the other is well understood. Things are more subtle when considering a general covariant system, where the further symmetry given by the invariance under diffeomorphism has to be implemented. 

In particular, the Hamiltonian operator is a constraint, whose non-polynomial structure results in a complicate action.  Furthermore, the operator needs to be regularized and this introduces quantization ambiguities.
In Loop Quantum Gravity there is an active research to define this operator (see for instance \cite{Giesel:2007wi}) but there is not a final agreement on its definitive form. The situation is much more developed in Loop Quantum Cosmology, where the symmetry reduction simplifies this study and a well-behaving Hamiltonian constraint is available \cite{Ashtekar:2006wn}.

On the other hand, the path integral approach has the advantage to overcome these complications and the ambiguities.  The quantum operators are replaced by $c-$numbers and the attention shifts to the meaning of the formula:
\be
   Z \sim \int Dg \ \ e^{\frac{i}{\hbar}\! \int \! \!R\sqrt{g}\,d^4\!x}.
   \label{ZZZ}
\ee 

A striking interpretation of this integral as a  \emph{sum over geometries} was obtained by Ponzano and Regge in 1968 \cite{Ponzano:1968uq}
in the simplified context of 3-dimensional Euclidean spacetime. The generalization to a 4-dimensional Lorentzian theory is due to Barrett and Crane \cite{Barrett:1997gw,
Barrett:1999qw} and has recently evolved into the present Spinfoam theory \cite{Rovelli:2011eq,Rovelli:2010wq,Rovelli:2010vv}.
%
%\mbox{(For an introduction, see for example \cite{Rovelli:2011eq}.) }
A \emph{spinfoam} is a 4-dimensional simplicial 2-complex $\cal C$
colored with \emph{spins} $j_f$ and intertwiners $i_e$, where
the labels stand respectively %for the vertices $v$ i.e. the 4-cell of $\cal C$,
the faces $f$ and the edges $e$ i.e. the 3-cells. This is the object that encodes the quantum geometry (Penrose's spin-geometry theorem).
In Loop Quantum Gravity the transition amplitudes are obtained by summing over all the possible spinfoams. %$\sum_\sigma$,
This corresponds to take the sum over the coloring $j_f$ and $i_e$,
the product of ``face amplitude'' $\prod_f d(j_f) $ and a product of vertex amplitudes $\prod_v A_v(j_e,i_v)$, that reads \cite{Engle:2007uq,Livine:2007vk,Engle:2007qf,Freidel:2007py,Engle:2007wy}\be
   Z=\sum_{j_f,i_e} \prod_f d(j_f) \ \prod_v A_v(j_e,i_v).
   \label{ZEPRL}
\ee 
%The form of this amplitude satisfies the fundamental requirements people were looking for.
%
%The amplitude of the individual histories is local, in the sense of being the product of face and vertex amplitudes.
%Every elementary vertex amplitude $A_v$ is expected to correspond to local elementary process, as usual in quantum field theory%
%\footnote{For an analogy, think for example to the vertex in QED vertex: a vertex is the elementary dynamical process that gives an amplitude to the boundary Hilbert space of two electrons and one photon.}.
%
%It is locally Lorentz invariant at each vertex, namely if we perform a local Lorentz transformation, the amplitude does not change.

Loop Quantum Gravity predicts a discrete spectrum for area and volume, therefore \eqref{ZEPRL} is ultraviolet finite since there are no transplanckian degrees of freedom. The infrared behavior can be kept under control by considering the quantum-deformed version of this amplitude \cite{Fairbairn:2010cp,Han:2010pz,Han:2010uq} that is IR finite. The q-deformed model has the further propriety to describe the cosmological constant coupling \cite{Ding:2011fk}.
The finiteness of \eqref{ZEPRL} is also preserved if fermions and Yang Mills fields are coupled \cite{Bianchi:2010bn}.
Finally,  \eqref{ZEPRL} is a discretization of the path integral for quantum gravity
that approaches  \eqref{ZZZ}  when taking the semiclassical limit in a suitable way\footnote{More precisely $A_v(j_e,i_v)$ approaches the exponential of the Regge action, which in turns approaches the action of GR.}.

\section{The spinfoam approach to cosmology}

Given the full spinfoam amplitude, this can be evaluated  for a homogeneous isotropic geometry described by a finite number of degrees of freedom. One can therefore study quantum cosmology directly from the spinfoam formalism by associating a cosmological interpretation to this amplitude \cite{Bianchi:2010zs}.%

This can be obtained by exploiting the characteristic features of the theory. These are slightly different with the ones that we are used, because of the effects of the invariance under diffeomorphism. 
In particular, this allows to define the Hilbert space of the theory as (the limit for refinements of) a graph space. The graph is analogue to the lattice used in QCD, but now the lattice is a ``coordinate" lattice where coordinates are gauges: the consequences are that taking the continuous limit there is no fixed point and no physical spacing of the lattice, therefore the lattice itself drops down. The concrete contribution of the graph is to provide a natural cut off of the infinite degrees of freedom of General Relativity, in the same manner that a finite number of particle in the infinite Fock space does in Quantum Field Theory.
This is particularly meaningful in order to define a cosmological system.
 The graph truncation can be seen as an expansion in modes, where the full theory can be recovered adding this modes one by one starting from the lowest, i.e. the cosmological one. In fact the cosmological principle allows as to approximate the universe by describing it with just one degree of freedom (the scale factor) or few ones. In spinfoam cosmology, truncating the theory to a given graph choses how many degrees of freedom we want to describe.

To fix the ideas, suppose to take a graph that corresponds to a triangulated 3-sphere and to compute the transition amplitude between a 3-sphere universe and a bigger 3-spere universe.
These two graph has to be thought as the support of \emph{boundary states} \cite{Oeckl:2003vu} and the amplitude we compute is a function of them. Diff-invariance dictates the need of boundary states, without whom we could not extract physical information from \eqref{ZEPRL} inserting bulk operator, as usually done in QFT.

Since we are interested in the semiclassical behavior of \eqref{ZEPRL}. we can use semiclassical states. In Loop Quantum Gravity, these are coherent states valued in $SU(2)$ \cite{Bianchi:2009ky,Flori:2008nw,Thiemann:2002vj,Thiemann:2000aa,Sahlmann:2001nv,Ashtekar:1994nx}. They have a nice geometrical interpretation in terms of the cellular decomposition dual to the graph considered, and in particular they carry the information about the (intrinsic and extrinsic) curvature and the area.
Coherent states provides a further tool to do cosmology.
In fact, it is possible to choose to peak them on a homogeneous and isotropic geometry, in a way that recovers the unique degree of freedom of the %Friedmann-Lemaitre-Robertson-Walker  
FLRW universe \cite{Magliaro:2010fk}. Since these states are realized as superposition of states on the boundary of the spinfoam, they naturally bear all the quantum fluctuations around the geometry described.

%\subsubsection{Transition amplitude between an initial and a final universe}

Given two boundary states of the (geometry of) the universe, we can compute the transition amplitude associated to them. In other words, we associate a dynamics to the kinematical states. This is given trough a vertex expansion, as usual in QFT.
We assume that the first order would dominate the expansion\footnote{The further orders have been started to be studied only very recently in \cite{Rovelli:2011kf}.}. At the first order, it is possible to study separately the probability of each state to arise from nothing (or to annihilate): this is exactly the same situation produced for the Hawking's  \emph{no-boundary} quantum states. This vertex amplitude, in the base of the coherent states of the geometry \cite{Bianchi:2010ys}, in the large-spin limit takes the form \cite{Vidotto:2011qa}:
%\vspace{-1mm}
\be\label{www}
~W_v(z) \sim
\sum_{j} \prod_{\ell}  \!{(2j_\ell+1)} 
\exp[-2t\hbar j_\ell(j_\ell+1)    - i  z j_\ell -i\lambda {\rm v}_o j^{\frac32}
]~.
\ee
This expression is a function of a unique complex variable $z$. The real part of $z$ is proportional to the scale factor squared and the imaginary part to the extrinsic curvature. The $t$ that appears in the exponential comes from the coherent states and represents their spread. The dependence on the graph for which the amplitude is calculated, reflected in the product over the $\ell$ links of the graph, can be reabsorbed once \eqref{www} is normalized. Finally, ${\rm v}_o j^{\frac32}$ is the volume associated to each node of the graph, and $\lambda$ is a parameter that yields the cosmological constant.

%	The evaluation of \eqref{www} at the saddle point shows that the amplitude is suppressed everywhere but where certain conditions on the real and the imaginary part of $z$ are satisfied. These can be combined to give
%	%
%	\be \frac{Re(z)^2}{Im(z)} =\frac{\lambda^2 {\mathtt{v}}_o^2}{4t\hbar}
%	\vspace{-1mm}\ee
%	that is, putting all together, the usual Friedmann equation for de Sitter space: 
%	%
%	$ \left({\dot{a}}/{a}\right)^2={\Lambda}/{3}
%	$.
%
The the transition amplitude obtained in this way results to be peaked on the correct classical solution, that satisfy the Friedmann equations.
This has been proven non only for flat space, but also in presence of a cosmological constant \cite{Bianchi:2011ym}, finding back correctly De Sitter space in the classical limit.
This results shows that it is possible to derive the Friedmann equation from the full covariant loop gravity, and give a further proof of the robustness of the vertex used, since it gives a good semiclassical limit.

This calculation can also provide further understanding about the relation between the covariant and the canonical formalism.
The transition amplitude can in fact be seen as a projector on the physical states. These are the states that satisfy the Hamiltonian constraint, therefore the transition amplitude should be annihilated by the Hamiltonin constraint.
We find that the transition amplitude obtained is in fact annihilated by a certain operator. In the large-distance approximation, where area on the cellular faces is larger than the Planck scale, we find that the classical expression for this operator turns out to be exactly
the usual Hamiltonian constraint for the FLRW dynamics.

%%%%%%%%%%%%%

\section{A path integral for Loop Cosmology}

The relation between the canonical and the spinfoam languages can be studied in the simplified context of cosmology. The quantum dynamics of cosmology is well under control in the canonical loop quantization 
\cite{Ashtekar:2006wn}
%\cite{Ashtekar:2003hd,Ashtekar:2006wn,Ashtekar:2007em,Ashtekar:2009vc} 
because of the simplification coming from the symmetry reduction of homogeneity and isotropy.

The idea \cite{Ashtekar:2009dn}  is to study \eqref{ZZZ} by inserting the simple Hamiltonian $ {\cal C}_H$ of Loop Quantum Cosmology:

\be\label{ACH}
 \langle \vec{\lambda}_F | \vec{\lambda}_0 \rangle_{\rm phy} = \frac{1}{2\pi} 
\int_{-\infty}^{\infty}
 {\rm d} \alpha \langle \vec{\lambda}_N | e^{i\alpha {\cal C}_H}
| \vec{\lambda}_0 \rangle  ~.
\ee
where $\alpha$ is a \emph{group averaging} parameter \cite{Higuchi:1991aa,Hartle:1997dc,Kaminski:2009qb}.
  The states on which the Hamiltonian constraint acts are quantum states of the geometry simply labelled by the total volume of the universe. The action of the Hamiltonian constraint derived from loop quantization is different from the usual Wheeler-deWitt expression: the operator is not a differential operator, but a difference operator. When it acts on the volume states, it changes them by discrete steps. This crucial peculiarity allows to treat the transition amplitude preserving the full general covariance\footnote{Notice that this was not the case for the treatment of the "wave function of the universe" of Hartle and Hawking and the works that followed (see for instance \cite{Halliwell:1984eu}).}.

This has been done for the flat %Friedmann-Lema\^itre-Robertson-Walker (
FLRW \cite{Ashtekar:2010ve}
%) 
and Bianchi type I \cite{Campiglia:2010jw} cosmologies. 
% controllare refrenze:
%\cite{Ashtekar:2009dn,Rovelli:2009kx,Ashtekar:2010ve,Campiglia:2010jw,Ashtekar:2010fk,Henderson:2010qd,Calcagni:2010ad}
In the first derivation of this path integral, a massless scalar field was included in the model for a double reason . First, it allows the theory to be straightforwardly deparametrized by treating it as an internal clock variable.  Second, it acts as a regulator since the implicit integration over it turns distributional transition amplitudes into regular functions. 
This procedure, however, leads to spinfoam amplitudes that are nonlocal in time.  A nonlocal spinfoam expansion can still be an effective computational 
tool, but does not match the structure of the spinfoam expression of the general theory \eqref{ZEPRL}. In the spinfoam amplitude locality is a foundational principle and full covariance under the choice of clock-time variables is strictly implemented. 

For this reason the construction of this path integral has been studied maintaining the full covariance under the choice of clock-time variables, namely without deparametrizing the theory. The distributional aspect of the transition amplitudes can be kept under control by introducing a  a regulator $\de$. 
The physical inner product obtained in this way is accurate up to some small error which vanishes as $\de\to0$.
The idea of such a regulator was introduced in \cite{Rovelli:2009tp} for the case where the spectrum of the eigenvalues of the Hamiltonian constraint operator is discrete, but the regulator used  there is not appropriate in  the continuous case. Notice that the choice of the regulator is not trivial: in fact, it has been shown that not all of them leads to \emph{local} amplitudes \cite{Henderson:2010qd}.

The regularized \eqref{ACH} takes the form
\be
 \lg \vl_F | \vl_0 \rg_{\pm\delta} = \f{i}{2\pi} ~ \sum_{M=0}^\infty \ 
\sum_{\vl_{1}...\vl_{M\!-\!1}}
\  \prod_{f}  A_f(\vl_{f})\ 
\prod_{v} A_v(\vl_{f})
\ee
where each term can be put in correspondence with the ones in \eqref{ZEPRL}. The transitions can be identified with the vertices of the spinfoam, therefore the a sum over the number of transitions that is equivalent to the sum over two-complexes; the sequences of steps without transitions can be identified with the faces of the spinfoam, the?sum over the values of the volume $\lambda_i$ is equivalent to the sum over colorings. This one-to-one correspondence shows that is possible to connect canonical and covariant formalism, at least in the simplified context of cosmology.

This approach is different respect with Spinfoam Cosmology, but the two approaches should hopefully converge and help to understand how the standard Loop Quantum Cosmology can be embedded in the full theory.

\section{Comments and further developments}

The quest for a fully general covariant path integral beyond formal expressions is essential in cosmology for the study of the early universe and in particular for the understanding of structure formation.

Spinfoam cosmology provides a viable framework to do concrete calculation in this context. Consider the boundary states chosen: these are homogeneous and isotropic, but they naturally includes quantum fluctuations. %In particular this provides a tool to study how quantum fluctuations of the geometry affect structure formation beyond usual perturbation theory.
Moreover their Hilbert space contains more degrees of freedom than the isotropic and homogeneous one \cite{Rovelli:2008ys,Battisti:2010kl}: inhomogeneities and anisotropies can be therefore naturally taken into account. % in this framework.  
 
The study of the geometry of the early universe, at the so called \emph{Big Bounce}, is a main motivation for this research program. In order to achieve it, we need to  study the transition amplitude beyond the strong classical limit taken so far. Present investigations are now exploring how to do this in the spinor framework.

The lesson of loop quantization have striking effects not only starting from the full spinfoam formalism, but also and particularly starting from cosmology. Loop Quantum Cosmology has already provided beautiful results, in particular singularity resolution: it stands as the most promising realization of the idea of a non-singular bouncing universe. 
The applications of its Hamiltonian constraint to the old ideas of the "wave function of the universe" of Hartle and Hawking has been proven powerful: the path integral defined in this way is fully general covariant and match the expansion in the full spinfoam theory. Furthermore, this 
shows a possibility to connect the canonical and the covariant framework in Loop Quantum Gravity.

% There are several issues that can be explored. I would like to mention in particular the presence of ambiguities in the regularization of the Hamiltonian constraint. In LQC has been introduced an ``improved dynamics" \cite{Ashtekar:2006wn}%,Corichi:2008fu
%, where the regularization is fixed by requiring a good semiclassical behavior. This choice can be made more robust if the resulting Hamiltonian will match with the one obtained in the covariant framework \cite{Bianchi:2010uq}.

% Hopefully, this model can help to understand how the standard LQC can be embedded in the full theory.

\begin{acknowledgement}
I would like to thank Mauro Carfora, Annalisa Marzuoli and Carlo Rovelli for their support and their patience. A special thanks to all the %Quantum Gravity 
group in Marseille and in particular to Eugenio Bianchi, Elena Magliaro, Antonino Marcian\`o and Claudio Perini, who contributed to this reserch with valuable discussions and a warm friendship.
\end{acknowledgement}

\providecommand{\newblock}{}

\end{document}